\documentclass[10pt,twocolumn]{article}
\usepackage[margin=0.85in]{geometry}
\usepackage{booktabs}
\usepackage{graphicx}
\usepackage{amsmath,amssymb}
\usepackage{microtype}
\usepackage[hidelinks]{hyperref}
\usepackage{xcolor}
\usepackage{caption}
\title{\textbf{GALOSH: Blind, Training-Free Denoising of Raw Bayer and sRGB
Images by Parallel-Friendly Local Shrinkage}}
\author{Yoshiro Sato\thanks{A U.S. provisional patent application covering the
methods described herein has been filed (App.\ No.\ 64/058,343, 6 May 2026).}}
\date{}

\begin{document}
\maketitle

\begin{abstract}
Classical training-free denoisers such as BM3D and non-local means owe much of
their strength to \emph{search}: content-dependent block matching whose memory
traffic and data-dependent control flow parallelize poorly and preclude
fixed-latency implementations. Learned denoisers reach the highest quality,
but they need training data, degrade outside their training domain --- which
we also observe --- and carry per-pixel compute budgets that effectively
require a GPU. We present \textbf{GALOSH} (Generalized Anscombe LOcal
SHrinkage), a redesign of training-free denoising that removes the search
entirely and aims at multi-domain coverage, speed, and quality at once: a
blind per-image Poisson--Gaussian noise fit, a generalized Anscombe transform,
a two-pass local Walsh--Hadamard shrinkage of luminance, and a
luminance-guided local regression of chrominance --- two deliberately
different operators for the two perceptually different noise components
(luminance grain vs.\ chroma blotches), each with its own strength control.
Every stage is local, data-independent, and regular --- the
same computation graph for every pixel of every image. One core serves two
domains: raw Bayer mosaics and sRGB/YUV images. On four real-noise benchmarks
(SIDD Medium and RawNIND, raw and sRGB) GALOSH is consistently the strongest
among the tested blind, training-free methods --- surpassing BM3D- and
NLM-family baselines even
when those are given an oracle noise level --- and approaches trained networks
on raw data while remaining below in-domain trained networks at high ISO in
sRGB. Being search-free makes it fast: $7\times$--$650\times$ faster than
the DL baselines on the same GPU at full benchmark size, depending on domain
(0.76\,s per 16-MP raw frame vs.\ 5.6\,s; 0.87\,s per 15.8-MP sRGB frame
vs.\ 13.5--565\,s), and the only strong method in the comparison that also
runs practically on plain CPUs.
The fixed, data-independent structure is designed to map naturally onto
fixed-point and streaming hardware, which we support with an operation-count
analysis and a working INT16 fixed-point realization.
\end{abstract}

\section{Introduction}
Two families dominate image denoising. \emph{Classical} training-free methods
--- BM3D~\cite{bm3d} and non-local means~\cite{nlm} --- need no data and
generalize across sensors, but they assume a known noise level and buy their
quality with content-dependent \emph{search}: block matching and neighborhood
weighting whose irregular memory access and data-dependent control flow resist
vectorization, GPU occupancy, and any fixed-latency budget. \emph{Learned}
methods (supervised or self-supervised CNNs and transformers) reach excellent
quality in the domain they were trained on, but need training pairs, carry
tens-of-thousands of MACs per pixel, and degrade --- sometimes catastrophically
(Sec.~\ref{sec:srgb}) --- outside their training distribution.

This paper asks: how much of the classical family's quality survives if the
search is removed \emph{by design}? GALOSH answers with a pipeline in which
every stage is local, regular, and data-independent: the noise model is
estimated blindly from the image itself; a generalized Anscombe transform
(GAT)~\cite{foi} makes a single global threshold valid everywhere; luminance is
denoised by two-pass local shrinkage on a Walsh--Hadamard decomposition (a
BayesShrink pilot~\cite{bayesshrink} followed by an empirical Wiener pass); and
chrominance is denoised by a luminance-guided local linear
regression~\cite{loess,kopf}. The luma/chroma split is deliberate: luminance
noise reads as grain and coexists with texture, whereas chrominance noise
reads as color blotches that are never acceptable, so the two paths use
different operators at different aggressiveness and expose independent
strength controls. No stage matches, searches, or branches on
content: the computation graph is fixed, so the method parallelizes trivially
and its cost is a constant per pixel.

The redesign is domain-general. The same core operates on two inputs:
\textbf{GALOSH-RAW} on the Bayer mosaic (where a $2{\times}2$ Walsh--Hadamard
transform of each quad provides the luma/chroma split) and
\textbf{GALOSH-YUV/RGB} on rendered sRGB images (where a linear-light BT.709
transform provides it). Our contributions:
\begin{itemize}\itemsep2pt
\item A blind, training-free, \emph{search-free} denoising pipeline whose
stages are all local, regular, and data-independent, with a shared core and two
domain front-ends (raw Bayer and sRGB/YUV) (Sec.~\ref{sec:method}).
\item Evidence across four real-noise benchmarks that GALOSH is the strongest
among the tested blind, training-free methods in both domains --- including
against
oracle-$\sigma$ classical baselines --- with an honest accounting of where
trained networks remain ahead (Sec.~\ref{sec:results}).
\item A speed study showing the practical value of search-freeness:
$7\times$--$650\times$ faster than the DL baselines on one GPU (domain-%
dependent), CPU-practical, and,
by an operation-count analysis plus a working INT16 realization, well suited
to fixed-point streaming implementations (Sec.~\ref{sec:speed},~\ref{sec:map}).
\end{itemize}

\section{Background}
GALOSH composes four classical ideas, each individually well established.
\emph{Variance stabilization}: Foi et al.~\cite{foi} showed that raw sensor
noise is accurately Poisson--Gaussian and can be fitted from a single frame;
the generalized Anscombe transform maps it to approximately unit-variance
Gaussian, and Makitalo--Foi~\cite{makitalo} give its unbiased exact inverse.
\emph{Transform-domain shrinkage}: BayesShrink~\cite{bayesshrink} sets a
near-optimal soft/hard threshold from the observed coefficient statistics, and
a second empirical Wiener pass with a pilot estimate is the standard refinement
(as in BM3D's two-step architecture~\cite{bm3d}).
\emph{Locally-weighted regression}: LOESS~\cite{loess} fits a low-degree local
polynomial under a kernel weight; the guided/joint-bilateral
family~\cite{kopf,guidedfilter} makes one channel steer another.
Beyond BM3D/NLM, the classical family includes low-rank patch models
(WNNM~\cite{wnnm}), patch-prior optimization (EPLL~\cite{epll}), learned
dictionaries (K-SVD~\cite{ksvd}), and the blind Noise Clinic
pipeline~\cite{noiseclinic}; these are likewise search- or optimization-heavy,
and we adopt BM3D/NLM --- with their CFA-aware and color variants --- as the
strongest widely-reproduced representatives in our comparisons.
GALOSH's claim is not any single component but their composition into a
search-free, fully-blind, two-domain pipeline --- replacing BM3D's block
matching with a cycle-spun local transform, and replacing its known-$\sigma$
assumption with a per-image blind fit whose accuracy the GAT converts into a
single valid global threshold.

\section{Method}\label{sec:method}
Figure layout: the shared core (Sec.~\ref{sec:core}) and the two domain
front-ends (Secs.~\ref{sec:raw},~\ref{sec:yuv}). All parameters are either
derived from the estimated noise model or fixed constants; there is no
per-image tuning.

\subsection{GALOSH core}\label{sec:core}
\paragraph{Blind noise estimation.}
A Poisson--Gaussian model $\operatorname{Var}[x]=\alpha\,\mathbb{E}[x]+\sigma^2$
is fitted from the noisy frame alone by robust regression of local
Laplacian-variance against local mean over overlapping blocks (a
Foi-style estimator~\cite{foi}), giving per-image $(\alpha,\sigma^2)$ with no
oracle and no calibration data.

\paragraph{Variance stabilization (GAT).}
The generalized Anscombe transform
$f(x)=\tfrac{2}{\alpha}\sqrt{\alpha x+\tfrac38\alpha^2+\sigma^2}$ maps the
signal-dependent noise to approximately unit variance, so one global threshold
is valid across the frame; a per-channel median-absolute-deviation (MAD) of a
high-pass residual supplies the residual scale. The inverse uses the exact
unbiased inversion~\cite{makitalo}.

\paragraph{Two-pass local luminance shrinkage (LOSH).}
Luminance is denoised by cycle-spun (stride-1) local shrinkage on overlapping
$8{\times}8$ Walsh--Hadamard blocks: pass~1 applies a robust
BayesShrink~\cite{bayesshrink} threshold to produce a pilot; pass~2 applies an
empirical Wiener gain computed from the pilot's coefficients. Overlapping
blocks are recombined by windowed overlap-add. This is BM3D's two-step
shrink-then-Wiener architecture with the collaborative (searched) 3-D stack
replaced by the dense cycle-spun local transform --- the step that removes all
content dependence.

\paragraph{Luminance-guided chrominance regression.}
Each chroma plane is denoised by a degree-1 locally-weighted regression on the
denoised luminance (a Y-guided LOESS~\cite{loess}; equivalently a guided
filter~\cite{guidedfilter} with a bilateral kernel~\cite{kopf}): within each
local window, kernel weights combine spatial distance and luminance
similarity, and a MAP-regularized slope/intercept regress chroma on luma. A
multi-scale residual pyramid extends the same operator to large-grain chroma
blotches. Because a degree-1 regression extrapolates at edges, every
reconstructed chroma value is clamped to the local input chroma range
(``the denoiser may not invent color the input lacks'').

\paragraph{Deliberate luma/chroma asymmetry.}
The luma/chroma split is not a data-format convenience; it is the perceptual
core of the design. Residual luminance noise reads as film-like grain and
coexists with texture, so the luma path shrinks conservatively in the
transform domain, preserving high-frequency structure. Chrominance noise reads
as color blotches that are never visually acceptable, so the chroma path
smooths far more aggressively --- safely, because the regression is anchored
to the denoised luma structure and clamped to the local input chroma range.
The two paths expose independent strength controls ($s_L$, $s_C$); every
benchmark in this paper uses the fixed defaults (no per-image tuning), but the
controls let a deployment trade grain retention against smoothness for each
component separately. This asymmetry is a major reason the perceptual margins
(LPIPS/DISTS) over the classical baselines are far larger than the PSNR
margins (Sec.~\ref{sec:results}). The classical color variants tacitly
acknowledge the same asymmetry --- CBM3D performs its block matching on the
luminance channel only, and OpenCV's colored NLM exposes separate
luminance/chroma strengths --- but both then apply the same search-based
operator to every component; GALOSH makes the asymmetry structural, with a
different operator per component.

\subsection{GALOSH-RAW}\label{sec:raw}
The raw front-end works directly on the Bayer mosaic in a single forward pass.
After the blind fit and GAT (core), a CFA-aware achromatic dark reference
(an IRLS estimate of the local achromatic floor) is subtracted; a
\emph{stride-1 cycle-spun} $2{\times}2$ Walsh--Hadamard transform converts each
Bayer quad into one luma plane $L$ (full resolution) and three chroma planes
$C_1,C_2,C_3$ (half resolution --- band-limit faithful, since the CFA samples
chroma at half the luma rate). $L$ receives the core two-pass LOSH; the chroma
planes receive the core guided regression at half resolution, and are returned
to full resolution by a luminance-guided joint-bilateral EWA jinc upsampling
(chroma edges snap to luma edges via the cross-channel prior; the jinc's
negative side lobes are anti-ringed by clamping to the local $2{\times}2$
input hull). Inverse WHT, dark restoration, and the exact inverse GAT yield a
denoised Bayer frame ready for any demosaicing back-end.

\subsection{GALOSH-YUV/RGB}\label{sec:yuv}
The second front-end accepts rendered sRGB images (the main API; a linear-RGB
entry point is trivial). Input is linearized by the inverse sRGB gamma and
converted to full-range BT.709 YCbCr; the noise on Y is fitted with the same
blind Poisson--Gaussian estimator (rendered noise remains signal-dependent
through the tone curve, which the GAT absorbs to first order). Y receives GAT
+ the core two-pass LOSH; Cb/Cr receive the core Y-guided regression at full
resolution (no WHT or upsampling is needed --- chroma is not subsampled in this
domain), with the same local chroma-range clamp reflected from the raw
pipeline. The output is clamped to $[0,1]$ and returned through the forward
transform chain. There is no separate chroma noise model: the regression
regularizer, held constant in the variance-stabilized space, plays that role.

\section{Experimental setup}\label{sec:exp}
\paragraph{Datasets.} Raw domain: \textbf{SIDD Medium}~\cite{sidd} (80
full-resolution smartphone Bayer images) and \textbf{RawNIND}~\cite{rawnind}
(1493 cross-camera $512^2$ real-noise raw crops). sRGB domain: the sRGB
renditions of the same two sets (SIDD's paired sRGB at full frame, $\sim$15.8\,MP; RawNIND
rendered to sRGB, all 1493 crops).

\paragraph{Protocol: everything blind.} Every method runs blind --- no
clean/noisy pairs, no per-image noise oracle. Classical baselines estimate
$\sigma$ from the input; GALOSH uses its own blind fit; DL baselines are their
published real-noise weights. Fidelity (PSNR/SSIM) is computed in each native
domain; perceptual metrics (LPIPS~\cite{lpips}, DISTS~\cite{dists},
NIQE~\cite{niqe}) on sRGB (raw results are rendered with per-image metadata).
On full-frame sRGB ($\sim$15.8\,MP), the DL baselines run tiled inference
(overlapping tiles, feathered), and LPIPS/DISTS are reported as the mean over
the same $1024^2$ tile grid for \emph{every} method --- a single uniform
protocol imposed by GPU-memory and metric-implementation limits at this
resolution, not a per-method choice.

\paragraph{Baselines are domain-appropriate by design.} Raw and sRGB denoising
are different tasks with different proper baselines, so the two domains use
different method sets: on raw Bayer, CFA-capable methods (BM3D-CFA, NLM-CFA,
their VST-front-end ablations, and the raw-trained Blind2Unblind~\cite{b2u});
on sRGB, color-image methods (CBM3D, colored NLM, a self-guided filter, and
the sRGB-trained NAFNet~\cite{nafnet}, Restormer~\cite{restormer},
SCUNet~\cite{scunet}). Note that the classical family itself adapts across
domains --- BM3D and NLM appear in both columns in domain-appropriate forms
(BM3D-CFA/NLM-CFA on the mosaic, CBM3D/Color-NLM on color images). This
adaptability is a virtue of the classical family, and GALOSH retains it ---
with one shared core behind thin domain front-ends --- while the trained
networks are bound to the domain and rendering of their training data.

\section{Results}\label{sec:results}

\subsection{Raw domain}\label{sec:rawres}
\begin{figure*}[t]
\centering
\includegraphics[width=\textwidth]{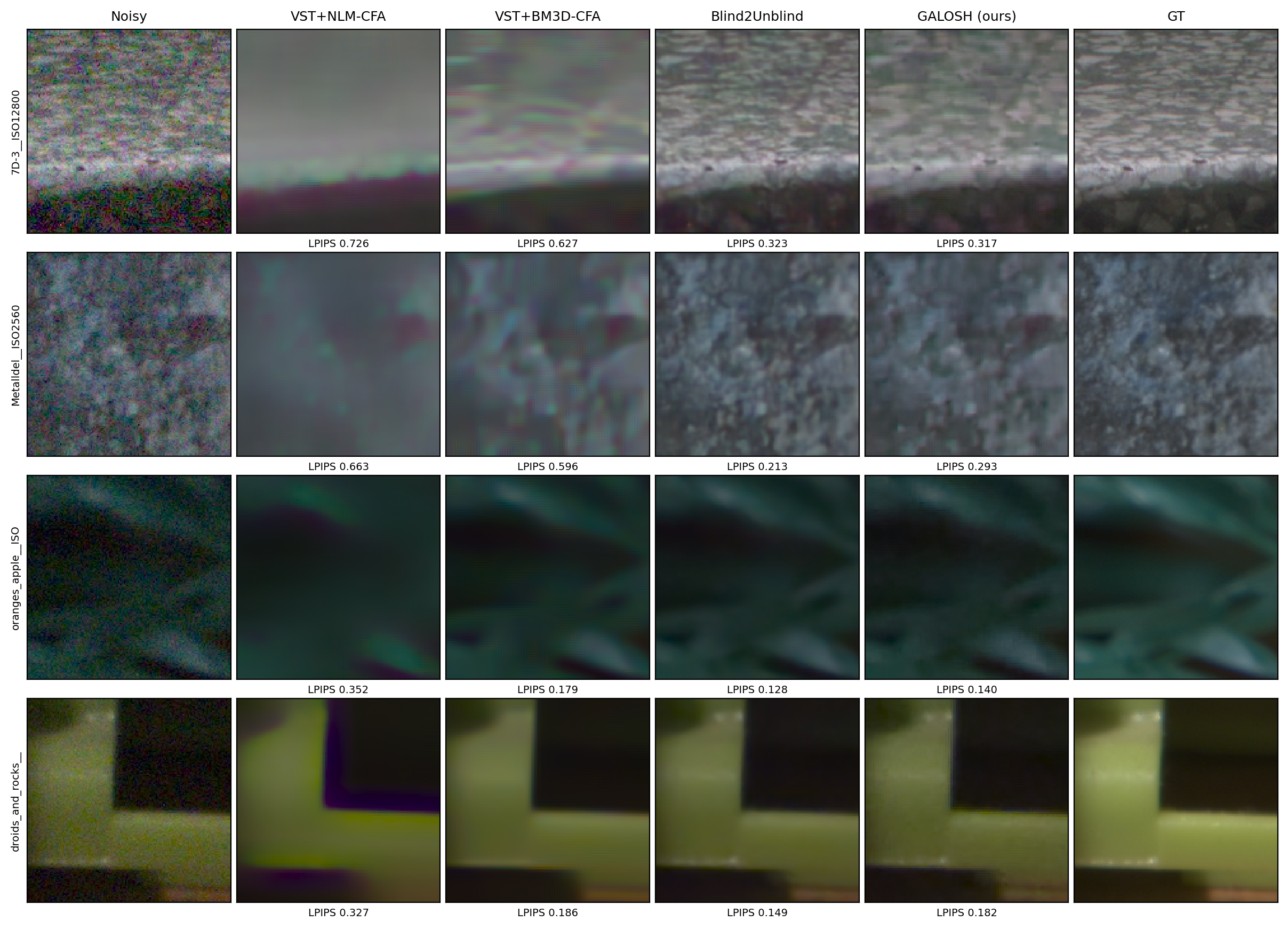}
\caption{Raw domain, RawNIND. Columns: Noisy / VST+NLM-CFA / VST+BM3D-CFA /
Blind2Unblind (trained) / \textbf{GALOSH (ours, blind)} / GT, with per-image
LPIPS ($\downarrow$). The classical baselines are given the noise-aware VST
front-end; GALOSH is fully blind.}
\label{fig:qual_rawnind}
\end{figure*}

\begin{figure*}[t]
\centering
\includegraphics[width=\textwidth]{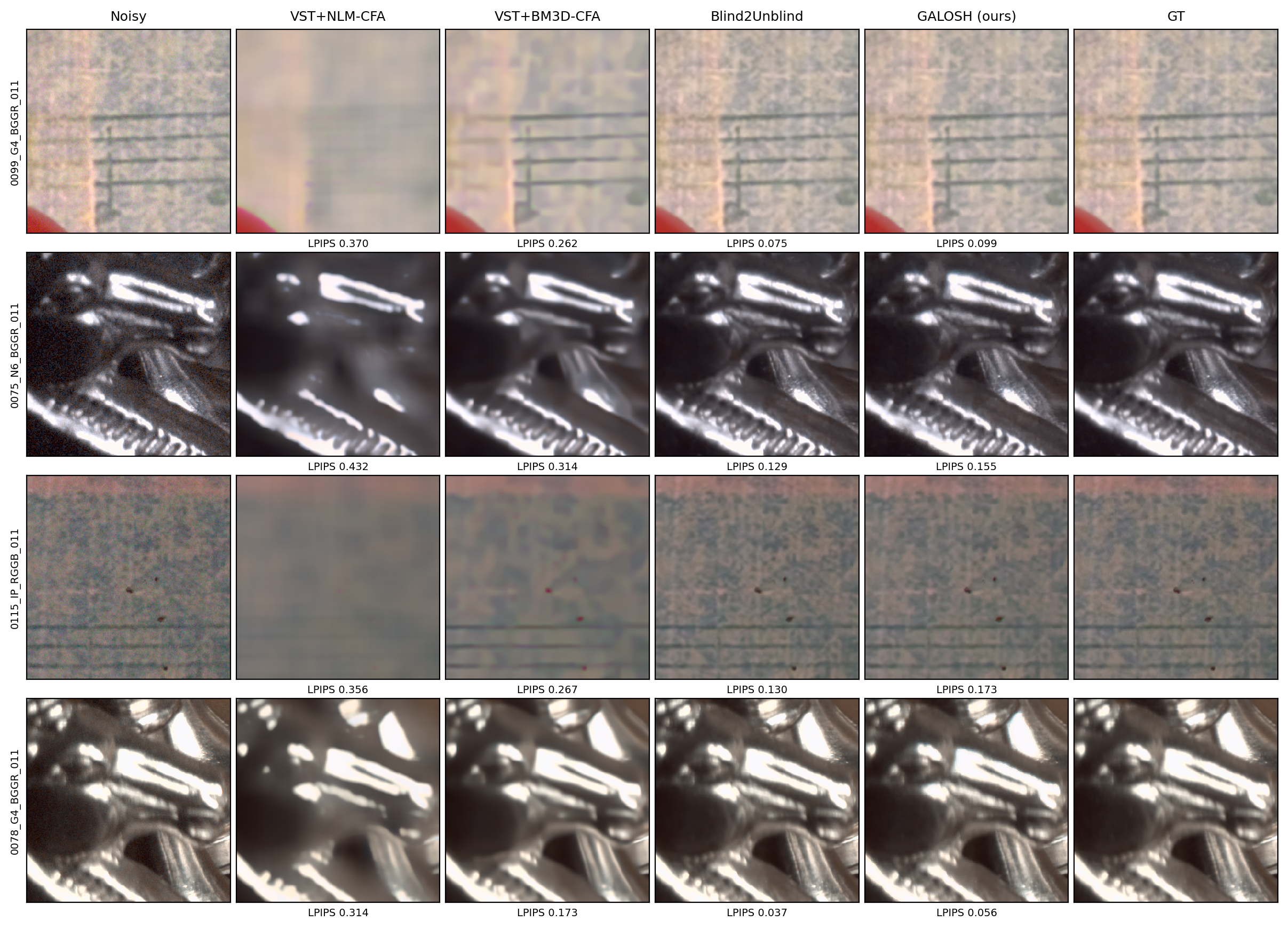}
\caption{Raw domain, SIDD Medium (same columns as Fig.~\ref{fig:qual_rawnind}).}
\label{fig:qual_sidd}
\end{figure*}

Tables~\ref{tab:sidd_medium} and~\ref{tab:rawnind} report the raw benchmarks.
Among blind, training-free methods GALOSH is the best on the perceptual
metrics on both datasets by a wide margin (LPIPS $0.203$/$0.240$ vs.\
$0.27$--$0.55$ for the BM3D/NLM family, with or without the shared VST
front-end), and competitive on PSNR/SSIM. Against the \emph{trained}
Blind2Unblind reference, GALOSH closes most of the perceptual gap on RawNIND
(LPIPS $0.240$ vs.\ $0.222$) without any training. The VST ablation rows
isolate the source of the advantage: giving the classical baselines GALOSH's
own variance-stabilizing front-end improves them only marginally --- the gap
is the search-free local shrinkage architecture, not the noise model alone.
Figures~\ref{fig:qual_rawnind} and~\ref{fig:qual_sidd} show representative
crops.

\subsection{sRGB domain}\label{sec:srgb}
\begin{figure*}[t]
\centering
\includegraphics[width=\textwidth]{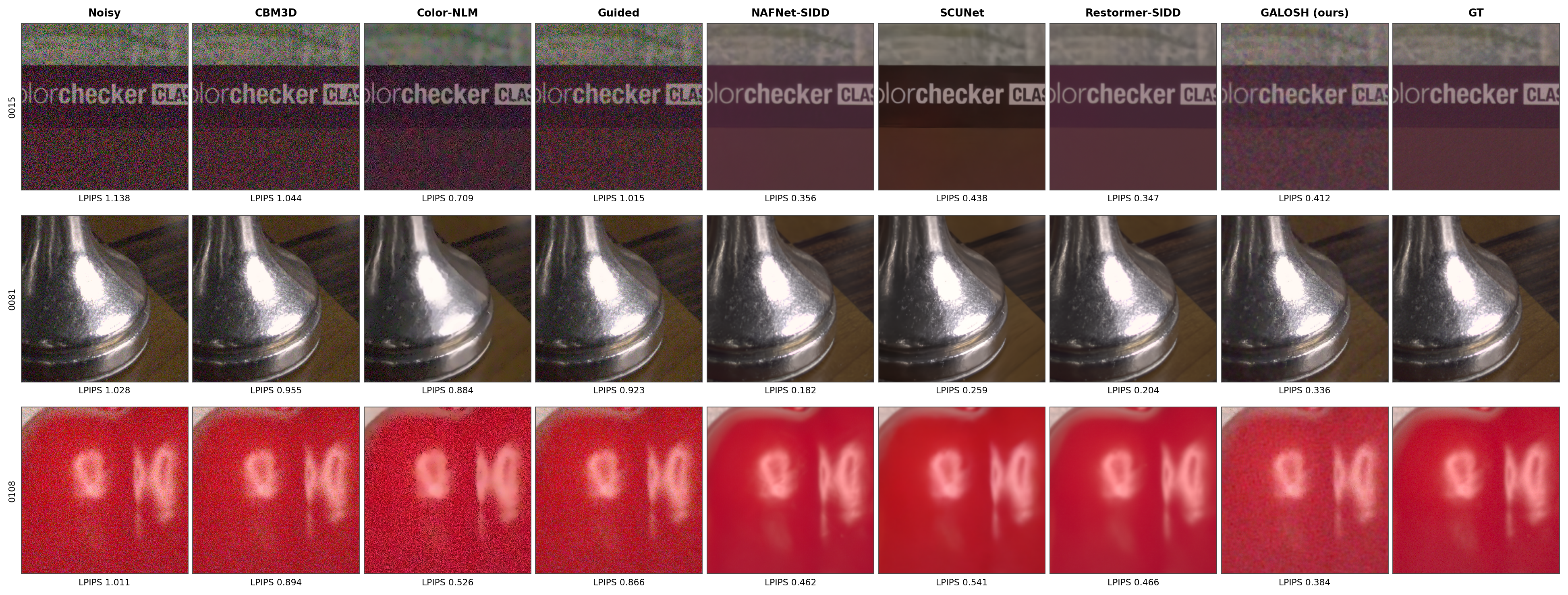}
\caption{sRGB domain, SIDD Medium sRGB. Columns: Noisy / CBM3D / Color-NLM /
Guided / NAFNet-SIDD / SCUNet / Restormer-SIDD / \textbf{GALOSH (ours)} / GT,
with per-image LPIPS. All methods blind; the classical baselines leave large
noise residuals on SIDD's correlated rendered noise, while GALOSH sits between
them and the SIDD-trained networks.}
\label{fig:qual_srgb_sidd}
\end{figure*}

\begin{figure*}[t]
\centering
\includegraphics[width=\textwidth]{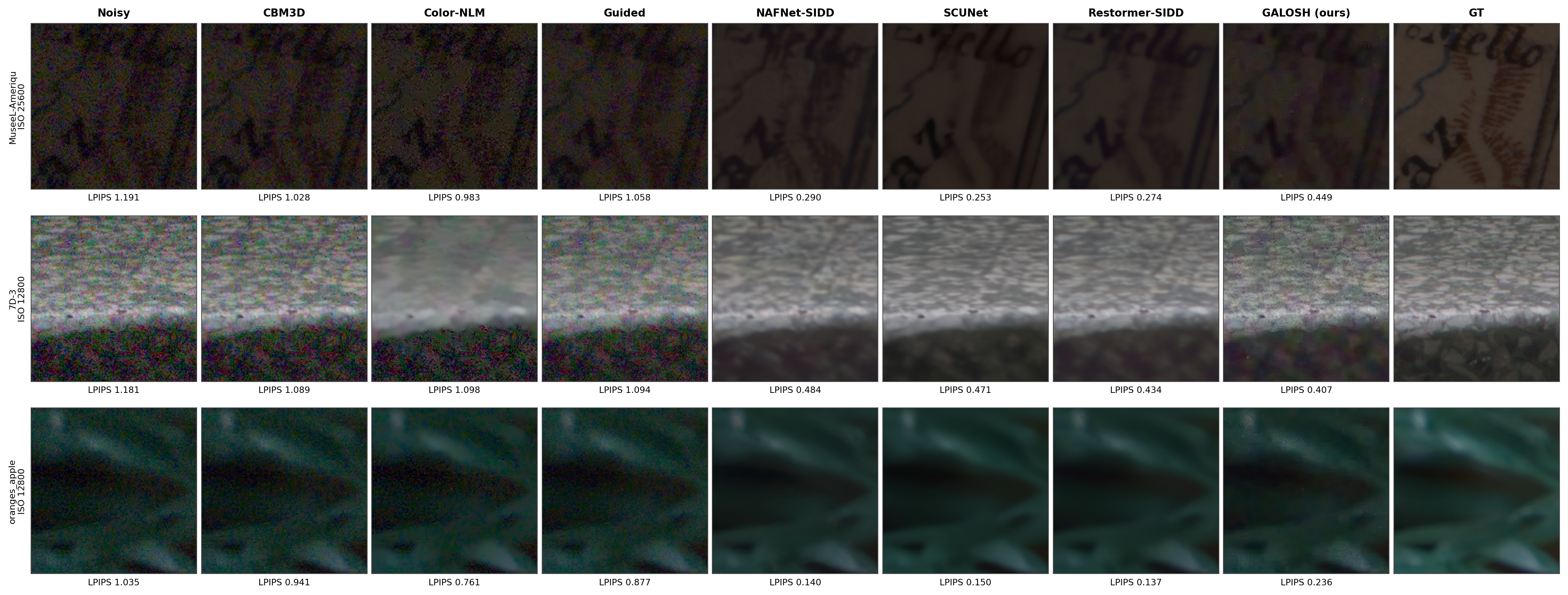}
\caption{sRGB domain, RawNIND (same columns as Fig.~\ref{fig:qual_srgb_sidd});
rows span ISO 12\,800--25\,600. Middle row: a case where GALOSH gives the best
LPIPS of all methods including the trained networks; top row: a dark
text scene where the trained networks lead --- the asymmetry discussed in
Sec.~\ref{sec:srgb}.}
\label{fig:qual_srgb_rawnind}
\end{figure*}

Tables~\ref{tab:sidd_srgb} and~\ref{tab:rawnind_srgb} report the sRGB
benchmarks; Figures~\ref{fig:qual_srgb_sidd} and~\ref{fig:qual_srgb_rawnind}
the corresponding crops. Three observations.

\textbf{(i) GALOSH decisively beats blind classical.} On SIDD sRGB the gap is
large (35.01\,dB vs.\ 27--29; LPIPS 0.314 vs.\ 0.53--0.72, at full frame): the spatially
correlated, signal-dependent rendered noise defeats both the baselines'
$\sigma$ estimators and their white-noise assumptions.

\textbf{(ii) The advantage is the algorithm, not the noise estimate.} On the
noisiest RawNIND subset (ISO\,$\geq$\,6400, $n{=}30$) we re-ran CBM3D and
Color-NLM with three $\sigma$ sources: their own blind estimate, GALOSH's
estimator, and the \emph{oracle} $\sigma$ computed from ground truth. Even
with the oracle, CBM3D reaches 21.33\,dB / 0.450 LPIPS and Color-NLM
21.26\,dB / 0.488 --- both below GALOSH's fully-blind 22.13\,dB / 0.413.
Handing the classical methods a perfect noise level does not close the gap.

\textbf{(iii) Honest asymmetry vs.\ trained DL.} On SIDD sRGB the
SIDD-trained networks are far ahead (NAFNet 41.94\,dB) --- expected, on their
training set. On cross-domain RawNIND the aggregate looks close
(Table~\ref{tab:rawnind_srgb}), but the aggregate is dominated by
already-clean low-ISO scenes; the per-ISO breakdown
(Table~\ref{tab:rawnind_srgb_iso}) is the honest view. At high ISO the
trained networks hold a moderate lead over GALOSH (+0.3--0.7\,dB), while
GALOSH in turn leads the classical family by $\sim$2\,dB and is the only
method in the blind training-free group that keeps gaining as noise grows.
Separately, NAFNet-SIDD diverges numerically on 37 near-black scenes (raw
outputs reach $\pm10^3$ before clipping); GALOSH and all other baselines remain stable on the same
inputs --- a robustness advantage of the fixed, training-free computation.

\subsection{Speed}\label{sec:speed}
All timings are per full benchmark image on one machine (RTX 4070\,Ti +
32-core CPU) and appear as the platform-split CPU/GPU columns of
Tables~\ref{tab:sidd_medium}--\ref{tab:rawnind}
and~\ref{tab:sidd_srgb}--\ref{tab:rawnind_srgb};
Table~\ref{tab:speed} adds a controlled same-input reference at 1080p.
Search-freeness pays directly. On the GPU, GALOSH is the fastest method in
every comparison: 0.76\,s per 16-MP raw frame vs.\ 5.6\,s for the trained
Blind2Unblind ($7\times$), and 0.87\,s per 15.8-MP sRGB frame vs.\ 13.5\,s
(NAFNet), 17.6\,s (SCUNet) and 565\,s (Restormer, VRAM-bound tiled attention)
--- $15$--$650\times$. On CPUs --- where the trained baselines are
impractical --- GALOSH's raw build is $2$--$3\times$ faster than BM3D-CFA and
NLM-CFA at full frame (Table~\ref{tab:sidd_medium}), and its 32-thread sRGB
build (2.5\,s per full frame) is faster than Color-NLM (2.8\,s) at far higher
quality. GALOSH is the
only method in the comparison that is simultaneously strong, GPU-fast, and
CPU-practical; at $512^2$ the GPU launch overhead dominates and the CPU build
is equally viable (Table~\ref{tab:rawnind}).

\section{Parallel and fixed-point mapping}\label{sec:map}
The property that removes the search also fixes the computation graph: every
pixel executes the same operations regardless of content. We quantify the
consequences without claiming a hardware implementation.

\paragraph{Operation count.} Instrumented over the full raw pipeline, GALOSH
costs $\approx\!3.4$k multiply-accumulates and $\approx\!140$ LUT/special
operations per pixel, resolution-independent ($3391$ at 1080p vs.\ $3370$ at
4K) --- $3.9$--$6.4\times$ below BM3D/NLM and $18$--$360\times$ below the
learned baselines, with local, data-independent access patterns throughout.

\paragraph{Fixed-point realization.} The shrinkage core admits a pure INT16
storage / INT32 accumulate fixed-point form. At INT32 storage, the GPU
streaming implementation is bit-exact against the INT32 CPU reference
end-to-end (verified on SIDD and RawNIND full frames); narrowing the line
buffers to the INT16 storage formats (luma Q10.5, chroma Q6.9) leaves the two
near-lossless ($\approx$58--65\,dB PSNR on full frames --- the residual is
exactly this storage quantization). Quality --- measured directly on the GPU
INT16 output --- matches the FP32 reference to within 0.1--0.7\,dB
(Tables~\ref{tab:sidd_medium},~\ref{tab:rawnind}). No 64-bit arithmetic is
required anywhere in the shipping path (wide intermediates use paired-32-bit
patterns).

\paragraph{Feasibility, not product.} By a MAC-array throughput model, the
per-pixel operation count is compatible with 1080p--4K-class streaming under
a $0.5$--$1$\,TMAC/s fixed-function budget --- modest by current ISP
standards; on-chip state is bounded by
line buffers of the largest vertical kernel ($\sim$200\,KB at 1080p in INT16).
We present these as design properties of the algorithm ("designed to map
naturally onto fixed-point streaming hardware"); building and verifying such a
stage is future work.

\section{Discussion and limitations}
GALOSH does not beat trained networks in their own domain, and at high ISO in
sRGB the trained networks retain a moderate quality lead; our claim is not
state-of-the-art quality but the combination --- blind, training-free,
multi-domain, search-free-parallel, and fast --- at quality that surpasses the
tested members of the
classical family even under oracle conditions. PSNR on raw SIDD remains
slightly below the trained Blind2Unblind. The blind estimator under-estimates
strongly correlated rendered noise (as all high-pass estimators do), which the
sRGB results absorb but do not eliminate. A learned lightweight corrector for
the $\alpha$ estimate showed promise in preliminary experiments and is
deferred to follow-up work. Video (temporal) extension is natural --- the
computation is already per-frame constant --- but unevaluated here.

\section{Conclusion}
Removing the search from classical denoising --- and replacing it with a
variance-stabilized, cycle-spun local shrinkage plus a luminance-guided local
regression --- retains and in fact extends the classical family's quality
(beating it even at oracle noise levels) while making the computation local,
regular, data-independent, and therefore parallel and fast. One training-free
core handles both raw Bayer mosaics and rendered sRGB images, runs
$7\times$--$650\times$ faster than learned baselines on a GPU (domain-%
dependent), remains practical on
CPUs, and is designed to map naturally onto fixed-point streaming hardware.


\begin{table}[t]
\centering
\caption{Raw-domain blind denoising on \textbf{SIDD Medium} (80 full-resolution
real-noise Bayer images). All methods are \emph{blind} (no clean/noisy pairs, no
per-image noise oracle) and \emph{training-free} except Blind2Unblind. Bold /
\underline{underline} = best / 2nd-best among the blind, training-free methods.
$\uparrow$: higher is better; $\downarrow$: lower is better.}
\label{tab:sidd_medium}
\setlength{\tabcolsep}{2pt}
\footnotesize
\resizebox{\columnwidth}{!}{%
\begin{tabular}{l ccccc rr}
\toprule
 & & & & & & \multicolumn{2}{c}{Time\,(s)}\\
\cmidrule(lr){7-8}
Method & PSNR$\uparrow$ & SSIM$\uparrow$ & LPIPS$\downarrow$ & DISTS$\downarrow$ & NIQE$\downarrow$ & CPU & GPU\\
\midrule
\multicolumn{8}{l}{\textit{GALOSH (ours) --- blind, training-free}}\\
\quad GALOSH FP32          & \textbf{48.13} & \textbf{0.9883} & \textbf{0.2030} & \textbf{0.1690} & 8.62 & 12.6 & 0.76\\
\quad GALOSH INT16\,$^{\dagger}$ & \underline{47.42} & \underline{0.9881} & \underline{0.2030} & \underline{0.1694} & \underline{8.62} & 103.6$^{\dagger}$ & 2.7\\
\addlinespace[2pt]
\multicolumn{8}{l}{\textit{Ablation --- shared VST front-end (no native estimation)}}\\
\quad VST\,+\,GALOSH core   & 44.35 & 0.9765 & 0.2132 & 0.1761 & \textbf{8.57} & 15.3 & 0.74\\
\quad VST\,+\,BM3D-CFA      & 45.96 & 0.9851 & 0.2932 & 0.2778 & 11.36 & 40.5 & ---\\
\quad VST\,+\,NLM-CFA       & 40.36 & 0.9679 & 0.4048 & 0.4194 & 12.98 & --- & 1.6\\
\addlinespace[2pt]
\multicolumn{8}{l}{\textit{Classical --- blind, native MAD}}\\
\quad BM3D-CFA             & 46.74 & 0.9862 & 0.2725 & 0.2613 & 10.55 & 40.4 & ---\\
\quad NLM-CFA              & 42.32 & 0.9729 & 0.3791 & 0.3901 & 12.05 & 27.1 & 1.2\\
\midrule
\multicolumn{8}{l}{\textit{Trained DL (upper reference, not blind)}}\\
\quad Blind2Unblind        & 49.07 & 0.9924 & 0.1169 & 0.1281 & 9.43 & --- & 5.6\\
\bottomrule
\end{tabular}}

\vspace{2pt}
{\footnotesize Per-image full-frame times, split by platform so only like
compares with like; every CPU entry is a single-threaded implementation; ``---'' = not
measured on that platform (BM3D has no GPU port; Blind2Unblind requires
CUDA). NLM-CFA: GPU = CUDA implementation, CPU = the skimage reference at the
same patch/search parameters (means over 12 full frames; RawNIND CPU entries
are medians over a 100-scene sample). GALOSH appears in
both columns --- the cross-platform anchor. $^{\dagger}$INT16 quality is measured
on the streaming INT16 GPU pipeline; its CPU time is the correctness-first
INT32 reference implementation (near-lossless to the INT16 output, not
speed-optimised) and its GPU time the streaming i16 implementation
(see Sec.~\ref{sec:map}).}
\end{table}

\begin{table}[t]
\centering
\caption{Raw-domain blind denoising on \textbf{RawNIND} (1493 cross-camera
$512{\times}512$ real-noise crops). Same protocol and highlighting as
Table~\ref{tab:sidd_medium}.}
\label{tab:rawnind}
\setlength{\tabcolsep}{2pt}
\footnotesize
\resizebox{\columnwidth}{!}{%
\begin{tabular}{l ccccc rr}
\toprule
 & & & & & & \multicolumn{2}{c}{Time\,(s)}\\
\cmidrule(lr){7-8}
Method & PSNR$\uparrow$ & SSIM$\uparrow$ & LPIPS$\downarrow$ & DISTS$\downarrow$ & NIQE$\downarrow$ & CPU & GPU\\
\midrule
\multicolumn{8}{l}{\textit{GALOSH (ours) --- blind, training-free}}\\
\quad GALOSH FP32          & \underline{30.49} & \underline{0.7908} & \textbf{0.2395} & \textbf{0.2127} & \underline{8.64} & 0.54 & 0.45\\
\quad GALOSH INT16\,$^{\dagger}$ & 30.17 & 0.7860 & \underline{0.2453} & \underline{0.2161} & \textbf{8.63} & 2.32$^{\dagger}$ & 0.35\\
\addlinespace[2pt]
\multicolumn{8}{l}{\textit{Ablation --- shared VST front-end (no native estimation)}}\\
\quad VST\,+\,GALOSH core   & \textbf{30.63} & \textbf{0.8163} & 0.2481 & 0.2198 & 8.65 & 0.49 & 0.60\\
\quad VST\,+\,BM3D-CFA      & 30.07 & 0.7792 & 0.4320 & 0.3563 & 13.04 & 2.29 & ---\\
\quad VST\,+\,NLM-CFA       & 29.03 & 0.7649 & 0.5452 & 0.4636 & 14.57 & --- & 0.68\\
\addlinespace[2pt]
\multicolumn{8}{l}{\textit{Classical --- blind, native MAD}}\\
\quad BM3D-CFA             & 30.28 & 0.7848 & 0.4268 & 0.3461 & 12.15 & 2.30 & ---\\
\quad NLM-CFA              & 29.54 & 0.7735 & 0.5190 & 0.4364 & 13.80 & 0.58 & 0.76\\
\midrule
\multicolumn{8}{l}{\textit{Trained DL (upper reference, not blind)}}\\
\quad Blind2Unblind        & 30.68 & 0.7932 & 0.2215 & 0.2018 & 8.93 & --- & ---\\
\bottomrule
\end{tabular}}

\vspace{2pt}
{\footnotesize Platform split as in Table~\ref{tab:sidd_medium}; at $512^2$
the GPU launch/sync overhead dominates GALOSH, so its CPU build is competitive
here. $^{\dagger}$See Table~\ref{tab:sidd_medium}. Blind2Unblind timing
omitted (results loaded from a precomputed cache).}
\end{table}

\begin{table}[t]
\centering
\caption{\textbf{Controlled-resolution speed reference} at 1080p
(1920$\times$1080 Bayer) on CPU (one core) and GPU (NVIDIA RTX 4070\,Ti); the
\emph{primary} speed comparison is the full-size per-image CPU/GPU columns of
Tables~\ref{tab:sidd_medium}--\ref{tab:rawnind}. Wall-clock, warm; GALOSH GPU
is the kernel pipeline (one-time context+build $\sim$90\,ms amortised), the
others end-to-end.}
\label{tab:speed}
\setlength{\tabcolsep}{8pt}
\resizebox{\columnwidth}{!}{%
\begin{tabular}{l rr}
\toprule
Method & CPU (1 core) & GPU\\
\midrule
\multicolumn{3}{l}{\textit{GALOSH (ours) --- blind, training-free}}\\
\quad GALOSH FP32 (o / o32)        & 1.38 s & \textbf{18.2 ms}\\
\midrule
\multicolumn{3}{l}{\textit{Classical (blind)}}\\
\quad BM3D-CFA                     & 6.33 s & ---\\
\quad NLM-CFA                      & 3.21 s & 0.42 s\\
\midrule
\multicolumn{3}{l}{\textit{Trained DL}}\\
\quad Blind2Unblind                & ---$^{\dagger}$ & 0.74 s\\
\bottomrule
\end{tabular}}

\vspace{2pt}
{\footnotesize GALOSH FP32 on GPU scales to 46\,ms at 4K and 72\,ms at 5K ---
$\approx$23$\times$ NLM-CFA and $\approx$40$\times$ Blind2Unblind on the same
GPU --- and on CPU it is $2$--$5\times$ faster than the classical baselines
(1.38\,s vs.\ 3.21/6.33\,s): fastest on both platforms.
$^{\dagger}$The Blind2Unblind reference implementation requires CUDA and does
not run on CPU. The INT16 fixed-point variant (Sec.~\ref{sec:map})
targets fixed-point streaming hardware rather than CPU/GPU throughput, so it
is not a speed entry.}
\end{table}


\begin{table}[t]
\centering
\caption{sRGB-domain blind denoising on \textbf{SIDD Medium sRGB} (80 scenes,
\textbf{full frame}, $\sim$15.8\,MP). Every method runs on the full image (the
DL networks via overlapped $1024^2$-tile inference; LPIPS/DISTS = mean over a
$1024^2$ tile grid). Every method is \emph{blind}: the classical baselines
estimate $\sigma$ from the input (skimage \texttt{estimate\_sigma}); GALOSH uses
its own blind noise model; the DL networks are their published real-noise
weights. Bold / \underline{underline} = best / 2nd-best among the blind,
training-free methods; the trained DL are an upper reference (NAFNet and
Restormer are trained \emph{on} SIDD).}
\label{tab:sidd_srgb}
\setlength{\tabcolsep}{2pt}
\footnotesize
\resizebox{\columnwidth}{!}{%
\begin{tabular}{l ccccc rr}
\toprule
 & & & & & & \multicolumn{2}{c}{Time\,(s)}\\
\cmidrule(lr){7-8}
Method & PSNR$\uparrow$ & SSIM$\uparrow$ & LPIPS$\downarrow$ & DISTS$\downarrow$ & NIQE$\downarrow$ & CPU & GPU\\
\midrule
Noisy input                & 25.69 & 0.424 & 0.790 & 0.380 & 5.86 & --- & ---\\
\addlinespace[2pt]
\multicolumn{8}{l}{\textit{Blind, training-free}}\\
\quad \textbf{GALOSH-YUV (ours)} & \textbf{35.01} & \textbf{0.837} & \textbf{0.314} & \textbf{0.243} & \underline{5.41} & 2.50 & 0.87\\
\quad CBM3D                & 27.09 & 0.492 & 0.721 & 0.349 & 5.52 & 118.6$^{1t}$ & ---\\
\quad Color-NLM            & \underline{28.92} & \underline{0.656} & \underline{0.534} & \underline{0.290} & \textbf{4.57} & 2.79 & ---\\
\quad Guided filter        & 27.43 & 0.523 & 0.687 & 0.349 & 5.79 & 0.71 & ---\\
\midrule
\multicolumn{8}{l}{\textit{Trained DL (reference)}}\\
\quad NAFNet-SIDD          & 41.94 & 0.942 & 0.167 & 0.154 & 7.69 & --- & 13.5\\
\quad Restormer-SIDD       & 40.94 & 0.935 & 0.181 & 0.175 & 8.05 & --- & 565$^{v}$\\
\quad SCUNet-real          & 36.09 & 0.902 & 0.252 & 0.234 & 9.30 & --- & 17.6\\
\bottomrule
\end{tabular}}

\vspace{2pt}
{\footnotesize Per-image times for the full 15.8-MP frame, split by platform;
``---'' = no implementation on that platform (the classical baselines have no
GPU ports; the DL networks are impractical on CPU). DL times are fresh-process
warm runs (a long-lived process inflates them via GPU-memory aging).
$^{v}$Restormer's tiled attention is VRAM-bound at full frame on this 12-GB
GPU. GALOSH is the only method with both platforms: GPU = the kernel pipeline,
CPU = the 32-thread build. CPU rows use all 32 cores except $^{1t}$: the
reference CBM3D package is effectively single-threaded.}
\end{table}

\begin{table}[t]
\centering
\caption{sRGB-domain blind denoising on \textbf{RawNIND-rendered sRGB} (all
1493 cross-camera $512^2$ crops). Same protocol and highlighting as
Table~\ref{tab:sidd_srgb}. This set is dominated by low-ISO scenes (684/1493
below ISO\,2000), which compresses the aggregate scores toward the noisy input
--- see the per-ISO breakdown in Table~\ref{tab:rawnind_srgb_iso}.}
\label{tab:rawnind_srgb}
\setlength{\tabcolsep}{2pt}
\footnotesize
\resizebox{\columnwidth}{!}{%
\begin{tabular}{l ccccc rr}
\toprule
 & & & & & & \multicolumn{2}{c}{Time\,(s)}\\
\cmidrule(lr){7-8}
Method & PSNR$\uparrow$ & SSIM$\uparrow$ & LPIPS$\downarrow$ & DISTS$\downarrow$ & NIQE$\downarrow$ & CPU & GPU\\
\midrule
Noisy input                & 22.13 & 0.662 & 0.479 & 0.253 & 8.58 & --- & ---\\
\addlinespace[2pt]
\multicolumn{8}{l}{\textit{Blind, training-free}}\\
\quad \textbf{GALOSH-YUV (ours)} & \textbf{23.33} & \textbf{0.841} & \textbf{0.265} & \textbf{0.227} & \textbf{7.12} & 0.077 & 0.012\\
\quad CBM3D                & \underline{22.49} & 0.720 & 0.396 & \underline{0.228} & 6.33 & 2.60$^{1t}$ & ---\\
\quad Color-NLM            & 22.48 & \underline{0.761} & \underline{0.353} & 0.225 & \underline{7.82} & 0.19 & ---\\
\quad Guided filter        & 22.50 & 0.724 & 0.402 & 0.236 & 8.62 & 0.012 & ---\\
\midrule
\multicolumn{8}{l}{\textit{Trained DL (reference)}}\\
\quad NAFNet-SIDD          & 22.58$^{\dagger}$ & 0.832 & 0.279 & 0.237 & 10.13 & --- & 0.13\\
\quad Restormer-SIDD       & 23.63 & 0.873 & 0.207 & 0.206 & 9.46 & --- & 0.35\\
\quad SCUNet-real          & 23.42 & 0.861 & 0.250 & 0.247 & 10.35 & --- & 0.14\\
\bottomrule
\end{tabular}}

\vspace{2pt}
{\footnotesize Platform split and footnotes as in Table~\ref{tab:sidd_srgb}
(times at $512^2$). $^{\dagger}$NAFNet-SIDD numerically diverges (network
outputs reach $\pm10^3$ before clipping) on 37 near-black scenes of this set
(PSNR\,$<$\,5 on those);
excluding them its mean is 23.05. All other methods, including GALOSH, remain
stable on the same scenes. Reported as-is.}
\end{table}

\begin{table}[t]
\centering
\caption{RawNIND sRGB \textbf{per-ISO} mean PSNR (dB). At low ISO the scenes
are already near-clean and no method can gain much; at high ISO --- where
denoising matters --- GALOSH clearly outperforms the classical baselines while
the trained DL are moderately ahead of GALOSH. We report this breakdown because
the aggregate of Table~\ref{tab:rawnind_srgb} hides both effects.}
\label{tab:rawnind_srgb_iso}
\setlength{\tabcolsep}{5pt}
\resizebox{\columnwidth}{!}{%
\begin{tabular}{l cccc}
\toprule
 & ISO$<$400 & 400--2k & 2k--12.8k & $\geq$12.8k\\
Method & (290) & (394) & (374) & (435)\\
\midrule
Noisy input      & 23.60 & 24.68 & 21.57 & 19.32\\
\addlinespace[2pt]
\textbf{GALOSH-YUV (ours)} & \textbf{23.66} & \textbf{25.20} & \textbf{22.58} & \textbf{22.06}\\
CBM3D            & \textbf{23.66} & 24.89 & 21.85 & 20.10\\
\midrule
NAFNet-SIDD      & 19.97$^{\dagger}$ & 24.44 & 22.67 & 22.56\\
Restormer-SIDD   & 23.68 & 25.36 & 22.77 & 22.78\\
SCUNet-real      & 23.52 & 25.26 & 22.62 & 22.36\\
\bottomrule
\end{tabular}}

\vspace{2pt}
{\footnotesize Bold = best among the blind, training-free methods.
$^{\dagger}$the near-black divergence cases (Table~\ref{tab:rawnind_srgb})
mostly fall in this band (29 of 37; the remaining 8 are in the 400--2k band,
mildly depressing that entry as well).}
\end{table}

\end{document}